\documentclass[11pt,a4paper]{article}
\usepackage[hyperref]{emnlp2020}
\usepackage{times}
\usepackage{latexsym}
\usepackage{amsmath}
\usepackage{amssymb}
\usepackage{lastpage}
\usepackage{hyperref}

\usepackage{fancyhdr}

\usepackage{arydshln}
\usepackage{subfigure}

\usepackage{microtype}
\usepackage{graphicx}  
\usepackage{comment}

\aclfinalcopy 

\usepackage{url}

\title{Transformer-XL Based Music Generation with \\ Multiple Sequences of Time-valued Notes}

\author{Xianchao Wu$^1$,
  Chengyuan Wang$^2$,
  Qinying Lei$^3$\thanks{Work done when Chengyuan and Qinying were internship students in Microsoft. These three authors contributed equally to this paper: Chengyuan was the major developer and Qinying was the major evaluator and advisor, supervised by Xianchao.} \\
  \\
  {\rm $^1$A.I.\&Research, Microsoft Development Co., Ltd.} \\
  {\rm $^2$ University of Science and Technology of China} \\
  {\rm $^3$ Communication University of China} \\
  {\tt \normalsize \{wuxianchao, forxms2019\}@gmail.com, lqy3110@hotmail.com}}

\date{}

\begin{document}
\maketitle
\begin{abstract}

Current state-of-the-art AI based classical music creation algorithms such as Music Transformer are trained by employing single sequence of notes with time-shifts. The major drawback of absolute time interval expression is the difficulty of similarity computing of notes that share the same note value yet different tempos, in one or among MIDI files. In addition, the usage of single sequence restricts the model to separately and effectively learn music information such as harmony and rhythm. In this paper, we propose a framework with two novel methods to respectively track these two shortages, one is the construction of time-valued note sequences that liberate note values from tempos and the other is the separated usage of four sequences, namely, former note on to current note on, note on to note off, pitch, and velocity, for jointly training of four Transformer-XL networks. Through training on a 23-hour piano MIDI dataset, our framework generates significantly better and hour-level longer music than three state-of-the-art baselines, namely Music Transformer, DeepJ, and single sequence-based Transformer-XL, evaluated automatically and manually.

\end{abstract}

\section{Introduction \& Related Work}

Music, one of the most important inventions of human being, has an extremely hugh market with billion-level listeners. 
Music is connected with emotion expression and language-independent, making it being one of the most intuitive and simplest ways of human communication. 

Creating music from scratch is a challenging work even for professional composers. Remarkably, popular and classical music generation using \textit{deep learning} algorithms and \textit{large-scale} MIDI files have achieved promising results during recent years. Main stream approaches include modelling music note sequences by borrowing ideas from language modelling in natural language processing field. 

A melody and arrangement generation framework for pop music was proposed in \cite{xiaoiceband2018wxc}. First, a chord-based rhythm and melody cross-generation model, employing recurrent neural networks (RNNs) such as gated recurrent units (GRUs) \cite{cho-etal-2014-learning}, was used to generate melody with chord progressions. Then, a multi-instrument co-arrangement model using updated GRUs was designed for multi-task learning for multi-track music arrangement. 

DeepBach \cite{deepbach2017} was a graphical model aiming at modelling polyphonic music and specifically hymn-like pieces through employing RNNs. Later, DeepJ was proposed in \cite{deepj2018} for style-specific music generating. Biaxial long short term memory (LSTM) \cite{hochreiter1997long,biaxial-lstm2017} was trained using piano roll note representations, for three major classical periods (baroque, classical, and romantic). DeepJ is capable of composing music conditioned on a specific mixture of composer styles. We take this model as one of our baselines and compare the differences of data representation and music dynamic sequential learning.

Compared with RNNs such as LSTMs or GRUs, Transformer \cite{NIPS2017_7181_transformer}, a sequence model based on (multi-head) self-attention mechanism, is more parallelizable for both training and inferring, and more interpretable. Transformer has achieved compelling results in tasks that require maintaining long-range coherence, such as neural machine translation \cite{NIPS2017_7181_transformer}, pre-training language models \cite{devlin-etal-2019-bert}, text-to-speech synthesis \cite{DBLP:journals/corr/abs-1809-08895}, and speech recognition \cite{DBLP:journals/corr/abs-1904-11660}.

Employing Transformer with relative attention mechanism \cite{shaw-etal-2018-self}, Music Transformer \cite{music-transformer-2018} was proposed for generating piano music with long-term structure. As depicted in Figure \ref{figure:note_4tuple}, performance events with absolute time intervals were employed for note sequence representation. In musical composition and performance, relative timing directly expressed by note-values is critically important. 
That is, we prefer the model learns from \emph{music scores} written by composers, instead of \emph{performance events} played by real-world players. 

NoteTuple \cite{transformer-nade-2018} groups a note's attributes as one event. However, the  TIME\_SHIFT used in performance events still brings hard-coding to the target of relative timing. In this paper, we propose a further \emph{relative} note representation method that projects tempo information into note representation, resulting a 4-tuple time-valued note representation which includes former note on to current note on, current note on to note off, pitch and velocity.

In particular, we are interested in generating extremely long music of hour level. We select Transformer-XL \cite{dai-etal-2019-transformer} which models extremely long (language) sequences by segment-level recurrences and relative position encoding. Different from \cite{donahue2019lakhnes}'s multi-instrumental music generation with event-based representation using transformer-XL, our proposal is to duplicate one Transformer-XL into four independently and jointly trained modules, taking our 4-tuple time-valued note sequences as inputs. Thanks to these time-valued data representation and four Transformer-XL networks trained in a joint way, our framework generates significantly better and hour-level longer music than three state-of-the-art baselines including Music Transformer, DeepJ, and single sequence-based Transformer-XL, evaluated automatically and manually. We extremely generated a continuous 36-hour music, with a stable high-level of note-density\footnote{Our full-version generated MIDI examples can be found at {https://pan.baidu.com/s/1i8pE7jEuWuWZy1DhW6XeJg} with code ckhv; and {https://drive.google.com/file/d/1VRoKY-INJ\_x1bte8SdTzvLao05uaZlIq/view?usp=sharing}.}. 





\section{Data Representation}\label{section:data-representation}

\begin{figure}
    \centering
    \includegraphics[width=8.0cm]{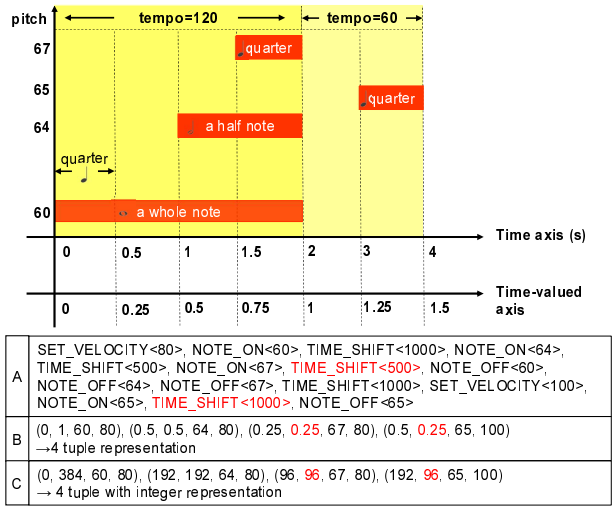}
    \caption{Illustration of our 4-tuple note representation.}
    \label{figure:note_4tuple}  
\end{figure}

We present a note by a 4-tuple $\langle on2on$, $on2off$, $pitch$, $velocity\rangle$. Here, $on2on$ and $on2off$ respectively stand for the \emph{note value} of from former note's on (start) to current note's on and from current note's on to its off (end). In addition, $pitch$ and $velocity$ are read from MIDI files directly using existing MIDI processing packages such as pretty\_midi\footnote{\url{http://craffel.github.io/pretty-midi/}} \cite{pretty-midi2014}. 

Figure \ref{figure:note_4tuple} illustrates the differences of between our 4-tuple note representation (B and C) and Music Transformer's \cite{music-transformer-2018} performance event representation (A). There is a velocity-80 $C$ Major chord arpeggiated with the sustain pedal active under tempo of 120 beats per minute (bpm). At the 2.0-second mark, the pedal is released, ending all of the three notes. At the 3-second mark, an $F$ quarter note is played at velocity 100 for 1 second under a tempo of 60 bpm. 

Suppose that the tempo for Figure \ref{figure:note_4tuple} is 120. The $on2on$ for the first note with pitch of 60 is 0. Its $on2off$ is computed by $2.0\times 120/(60\times4)=1.0$, where 2.0 is the two seconds that this note is on, 120 is the tempo, 60 for 60 seconds per minute, and 4 for the reciprocal of $1/4$ note (crotchet). Through this way, we list the 4-tuple representations of these four notes at B. In addition, as shown in C in Figure \ref{figure:note_4tuple}, we project these float representations of $on2on$ and $on2off$ into integers by producing them with 384=$128\times 3$. Note that 3 is introduced here for covering tritone cases. 

In our 23-hour experiment data, the maximum integer reaches to 3,840, as long as ten tritones. Generally, the major difference is that we use these integer-style time values instead of TIME\_SHIFT as employed in Music Transformer \cite{music-transformer-2018}. Our usage of time-valued notes aligns with note values natural usage in music book score. 

In addition, TIME\_SHIFT can possibly cause the confusion of time value information. That is, if we use the absolute time intervals to represent notes, then (1) one note can correspond to different time intervals and (2) one time interval can also correspond to different notes, under different tempo. The generated music is less user-friendly for composers' post-editing. Furthermore, in the performance event sequence, NOTE\_ON and NOTE\_OFF of various notes are mixed together which breaks the independences of notes and causes losing of time value information. Intuitively speaking, NOTE\_ON and NOTE\_OFF are alike brackets and should always be paired in one sequence for training. However, this is not guaranteed by event-length based sequence segmenting. Since the start and end time of each note are obtained by computing NOTE\_ON and NOTE\_OFF based on TIME\_SHIFT. As will be shown in our experiments, this computing process is problematic and frequently causes unstable rhythm (Figure \ref{figure:note-seq-compare}), especially during (10-minute and longer) long-time music generation. 

Piano roll representation of notes is used in DeepJ \cite{deepj2018} as a ``dense'' representation of MIDI music. A piece of music is a $N\times T$ binary matrix where $N$ is the number of playable notes and $T$ is the number of time steps. Considering the difference in holding a note versus replaying a note, note play and replay (i.e., when a note is re-attacked immediately after the note ends, with no time steps in between successive plays) are utilized jointly to define note representation. However, piano roll is not dense since there are so many zeros in the play/replay matrices: only a few notes are attacked during each time step and all others are zero. It is further not easy to employ piano roll representation for sequential learning for music notes' ``language modelling''. 

\section{Transformer-XL Training with Four Sequences of Time-valued Notes}\label{section:4transformer-xl}
\subsection{Transformer-XL}

Transformer-XL\footnote{{https://github.com/kimiyoung/transformer-xl}} \cite{dai-etal-2019-transformer} was designed to enable Transformer \cite{NIPS2017_7181_transformer} to learn dependency (of among language words or music notes in this paper) beyond a fixed length  without disrupting temporal coherence. It consists of two updates, a segment-level recurrence mechanism and a positional encoding scheme. These allow Transformer-XL not only capturing longer-term dependency but also resolving the language/music context fragmentation problem. Motivated by its outstanding performance in terms of language modelling, we adapt this framework for music generation through learning a ``language model'' by using time-valued \emph{notes} instead of words as the fundamental units.

Formally, let $\textbf{s}_{\tau-1}=[x_{\tau-1,1}$, $\cdots$, $x_{\tau-1,L}]$ be the $(\tau-1)$-th segment with length $L$ (e.g., $L$ notes in music and $L$ words in natural language), 
$\textbf{h}_{\tau-1}^{n-1}\in \mathbb{R}^{L\times d}$ is the $(n-1)$-th layer (e.g., in Transformer) hidden state sequence for $s_{\tau-1}$ in which $d$ is the hidden dimension. Then, for the next segment $\textbf{s}_{\tau}$, the hidden state of its $n$-th hidden layer is computed by:
\begin{align}
 \tilde{\textbf{h}}_{\tau}^{n-1} & = [\text{SG}(\textbf{h}_{\tau-1}^{n-1})\circ \textbf{h}_{\tau}^{n-1}]; \label{eq:h_recurrent}\\
 \textbf{q}_{\tau}^{n},\textbf{k}_{\tau}^{n},\textbf{v}_{\tau}^{n} & = \textbf{h}_{\tau}^{n-1}\textbf{W}_q^\top, \tilde{\textbf{h}}_{\tau}^{n-1}\textbf{W}_{k,(E)}^{\top}, \tilde{\textbf{h}}_{\tau}^{n-1}\textbf{W}_v^{\top}; \\
 \textbf{h}_{\tau}^{n} & = \text{Transformer-Layer}(\textbf{q}_{\tau}^{n},\textbf{k}_{\tau}^{n},\textbf{v}_{\tau}^{n}). \label{eq:transformer_layer} 
\end{align}
Here, the function $\text{SG}(\cdot)$ stands for stop gradient, i.e., the gradient of $\textbf{h}_{\tau-1}^{n-1}$ will not be updated basing on the next segments. The notation $[\textbf{h}_u\circ \textbf{h}_v]$ means the concatenation of two hidden sequences along the length dimension to extend the context. $\textbf{W}_{q,k,v}$ stand for trainable model parameters. Compared with Transformer \cite{NIPS2017_7181_transformer}, the major update here is the usage of $\textbf{h}_{\tau-1}^{n-1}$, previous segment's hidden state sequence of $(n-1)$-th layer, for computing an interim sequence $\tilde{\textbf{h}}_{\tau}^{n-1}$ and its further usage for computing the \textit{extended-context enhanced} sequences $\textbf{k}_{\tau}^{n}$ and $\textbf{v}_{\tau}^{n}$, to be retrieved by the query sequence $\textbf{q}_{\tau}^{n}$. 

Transformer-XL applies this recurrent mechanism as defined in these equations alike $\textbf{h}_{\tau}^{n}$=$\text{recurrent}(\textbf{h}_{\tau-1}^{n-1},\textbf{h}_{\tau}^{n-1})$ to every two consecutive segments of a corpus. This essentially creates a segment-level recurrence in the hidden states of various transformer layers. As a result, the contextual information is allowed to go way beyond just two segments.

In the standard Transformer, the attention score between a query $q_i=(\textbf{E}_{x_i}+\textbf{U}_i)$ (i.e., embedding vector $\textbf{E}_{x_i}$ adds with $i$-th \textit{absolute} position encoding $\textbf{U}_i$) and a key vector $k_j=(\textbf{E}_{x_j} + \textbf{U}_j)$ within the same segment can be decomposed as:
\begin{equation}
 \textbf{A}_{i,j}^{\text{abs}}=q_i^{\top}k_j=\left(
   \textbf{E}_{x_i}^{\top} + \textbf{U}_i^{\top}
 \right)\textbf{W}_q^{\top}\textbf{W}_k
 \left(
   \textbf{E}_{x_j} + \textbf{U}_j
 \right).
\end{equation}
The shortage of absolute position encoding $\textbf{U}$ is that it is not able to distinguish the difference of one same position appearing in different segments. Following the idea of relative positional encoding \cite{shaw-etal-2018-self}, a relative distance $\textbf{R}_{i-j}$ is introduced to describe the relative positional embedding between $q_i$ and $k_j$. Here, $\textbf{R}$ is a sinusoid encoding matrix alike the one used in \cite{NIPS2017_7181_transformer} without learnable parameters. The relative attention score is then computed by:
\begin{equation}
 \textbf{A}_{i,j}^{\text{rel}}=\left(
   \textbf{W}_q\textbf{E}_{x_i} + u:v
 \right)^{\top}
 \left(
   \textbf{W}_{k,E}\textbf{E}_{x_j} + \textbf{W}_{k,R}\textbf{R}_{i-j}
 \right).
\end{equation}
Two trainable vectors $u,v\in\mathbb{R}^d$ are used to replace $\textbf{U}_i\textbf{W}_q$ and are respectively used for multiplying with $\textbf{W}_{k,E}\textbf{E}_{x_j}$ and $\textbf{W}_{k,R}\textbf{R}_{i-j}$. In addition, $\textbf{W}_k$ is deliberately separated into two weight matrices $\textbf{W}_{k,E}$ and $\textbf{W}_{k,R}$ for respectively producing content-based and position-based key vectors.

Then, the $n$-th ($n$ ranges over 1 to $N$) Transformer-Layer used in Equation \ref{eq:transformer_layer} by employing relative position encoding mechanism is computed by:
\begin{align}
  \textbf{A}_{\tau,i,j}^{\text{rel},n}&=\left(
   {\textbf{q}_{\tau,i}^{n}} + u:v
 \right)^{\top}
 \left(
   \textbf{k}_{\tau,j}^{n} + \textbf{W}_{k,R}^{n}\textbf{R}_{i-j}
 \right); \\
 \textbf{a}_{\tau}^n & = \text{Masked-Softmax}(\textbf{A}_{\tau}^{\text{rel},n})\textbf{v}_{\tau}^n; \label{eq:masked-softmax}\\
 \textbf{o}_{\tau}^n & = \text{LayerNorm}(\text{Linear}(\textbf{a}_{\tau}^n)+\textbf{h}_{\tau}^{n-1}); \label{eq:add-norm-1}\\
 \textbf{b}_{\tau}^n & = \text{Positionwise-Feed-Forward}(\textbf{o}_{\tau}^n); \label{eq:feed-forward}\\
  \textbf{h}_{\tau}^n & = \text{LayerNorm}(\text{Linear}(\textbf{b}_{\tau}^n)+\textbf{b}_{\tau}^{n})\label{eq:add-norm-2}.
\end{align}

\subsection{Joint Training}

\begin{figure*}
    \centering
    \includegraphics[width=15cm]{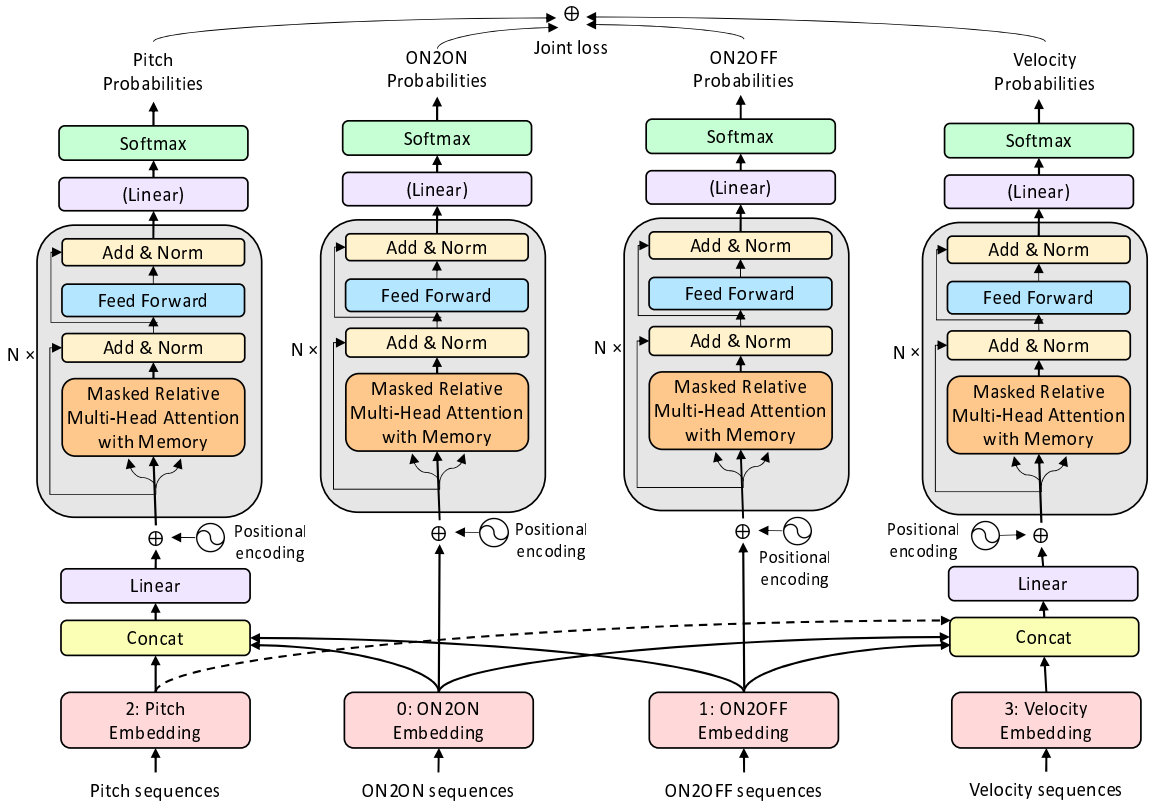}
    \caption{Our proposed framework with four transformer-xl networks.}
    \label{figure:4transformerxl}
\end{figure*}

Figure \ref{figure:4transformerxl} depicts our framework that leverages four Transformer-XL networks, 
corresponding to $on2on$, $on2off$, $pitch$, and $velocity$ in the 4-tuple time-valued notes. The first layer embeds these four sequences. 
Then, $on2on$ and $on2off$'s embedded sequences are sent to $pitch$ and $velocity$ for including of time value information of notes. In addition, $pitch$'s embedded sequences are sent to $velocity$ as well to provide an influence. Next, $pitch$ and $velocity$ concatenate their own embedded sequences with external sequences. The vector dimension of each position in the sequence will be thrice for $pitch$ and fourfold for $velocity$. 
We additionally employ a linear layer for dimension resizing before sending them to the memory sensitive blocks of Transformer-XL. Typically, each block mainly contain two components, a masked relative multi-head self-attention with memory (from former segment) and a position-wise feed forward layer. This block is repeated $N$ (e.g., 6 in the original Transformer \cite{NIPS2017_7181_transformer}) times. After these $N$ blocks for the four Transformer-XL, we employ a linear layer and softmax function to compute probabilities (i.e., normalized scores) of predicted sequences. We finally compute cross entropy loss for each sequence and add them up for the target loss optimizing. 

In particular, the ``masked relative multi-head attention with memory'' block is computed by Equation \ref{eq:h_recurrent} to \ref{eq:masked-softmax}. Then, a residual function is defined in Equation \ref{eq:add-norm-1} for ``add \& norm''. The next ``add \& norm'' residual layer is defined in Equation \ref{eq:add-norm-2} for the position-wise feed forward layer. 

In our proposed framework, the four sequences have relations in two places. First, concatenation of time valued $on2on$ and $on2off$ embedded sequences to $pitch$ and $velocity$. Second, joint loss which sums up the losses of the four sequences. The motivation of our designing is to both ensure the relative \textit{independence} of each sequence's development and the mixed \textit{influence} of from time-valued notes to pitch and velocity.

\section{Experiments \& Evaluations}\label{section:experiments}

\subsection{Data}

We collect 374 piano MIDI files from the web\footnote{\url{http://www.piano-midi.de/}}, which are hand-made from composers' \emph{piano music scores} with correct tempo information instead of players' performances\footnote{Such as the MAESTRO dataset in {https://magenta.tensorflow.org/datasets/maestro} with unchanged tempo information due to automatic transcribing from wav files to MIDI.} during periods of baroque, classical, and romantic. In the baroque period, Bach's fugue and prelude are mainly included. Classical period mainly contains products written by Beethoven, Mozart, Brahms, and Haydn. For the romantic period, we collect composers such as Chopin, Liszt, Mendelssohn, Schubert, and Tschaikovsky. Table \ref{table_data_midi} lists statistical information of our train, validation, and test sets.  

\begin{table} 
\begin{center}
\begin{tabular}{l|r|r|r}
\hline
 & train & validation & test \\
\hline\hline 
\# MIDI files & 299 & 34 & 41 \\ 
\# baroque & 69 & 8 & 10 \\ 
\# classical & 86 & 10 & 11 \\
\# romantic & 144 & 16 & 20 \\
length (hours) & 18.18 & 2.82 & 2.44 \\  
\hline
\end{tabular}
\end{center}
\caption{Statistical information of 3 data sets. }\label{table_data_midi}
\end{table} 

\subsection{Experiment Setups}

We perform our experiments under a NVIDIA P40 GPU card, with Tensorflow 1.13.1\footnote{\url{https://www.tensorflow.org/}} running under cuda 10.0 and cudnn 7.5.0. We implement our joint training framework (Figure \ref{figure:4transformerxl}) basing on Transformer-XL. The number of Transformer layers is 6, the number of heads is 8 with a dimension of 64. Embedding and hidden layer dimensions are identical to be 512. The dropout ratio is set to be 0.1. Memory length is 1,024. We use Adam algorithm \cite{kingma2014method} to optimize our networks. We retrain DeepJ\footnote{\url{https://github.com/calclavia/DeepJ}} with piano roll inputs, Music Transformer\footnote{{https://github.com/tensorflow/magenta/tree/master/ magenta/models/score2perf}} and single-sequence Transformer-XL both with performance event inputs, all using our 23-hour datasets. 

\subsection{Subjective Evaluations}

For human evaluations, we separate all participants into two groups, professional composers and non-composers. Participants in the professional group are people who hold education degrees in music composition or electronic music composition and production, including Central Conservatory of Music, Shanghai Conservatory of Music, Communication University of China, Eastman School of Music, Berklee School of Music, Peabody Institute of JHU, and Steinhardt of NYU. 

\subsubsection{Human or AI test}

Following \cite{deepbach2017}, we first design a ``Human or AI'' test for all participants. We prepared a mixed music collection with 5 music compositions composed by professional human composers and 11 pieces composed by our model for people to judge whether they are composed by humans or by AI. 39 professional composers were asked to rate for each music piece they heard from the music composition theory aspect while 61 non-composers were asked to rate by their subjective feelings. Our rating scale has five levels (points 1 to 5) including nonsense, basic, good, high-level, and masterpiece.

In order to avoid participants being affected by some other aspects such as musical instrument timbre and recording reverbs, all of the testing music whatever composed by human or AI are MIDI files and were exported from the same virtual piano (i.e., Logic's Steinway Grand Piano). Also, we limited the length of each music piece to be around 30 seconds, and every participant was asked to listen to 16 pieces. We limited the listening time to avoid auditory fatigue.

\begin{table} 
\begin{center}
\begin{tabular}{l|r|r|r|r}
\hline
 & It's Human & It's AI &  Avg. & Avg. \\
 & All (Pro.) & All (Pro.) & Pro. & All \\ 
 \hline
Human & 218 (98) & 282 (97) & \textbf{3.24} & 3.01 \\ 
AI & \textbf{411 (132)} & \textbf{689} (\textbf{300}) & 2.94 & \textbf{3.11} \\ 
\hline
\end{tabular}
\end{center}
\caption{Human vs. AI evaluation results. }\label{table_human_or_ai}
\end{table} 

The comparison results are listed in Table \ref{table_human_or_ai}. Among professional composers, the average human music compositions are scored higher than that of AI. However, among all participants, our AI music is scored  higher (3.11 vs 3.01) than that of the real human pieces which demonstrates that our AI music composition quality is reasonably close to the quality of human composers. Even some of the pieces transcends humans' work according to some single ratings. Interestingly, for all evaluators, human composers' music are judged to be AI-made in a dominate way.

\subsubsection{Pairwise Comparison}

Our second test is to compare the perceived sample quality with three baseline models. We carried out listening tests comparing the baseline Music Transformer (MusicT), DeepJ, and single sequence-based Transformer-XL (1-seq txl). Our model is denoted as Mtxl. Participants (15 composers and 30 non-composers) were presented with two musical excerpts that were generated from two different models but were given the same priming note. Then the participants were asked to rate which one they preferred more. We generated 7 samples each model with a different prime, and compared them to three other models. In addition, we asked every participant to rank the reason of choice from four musical aspects: melody, harmony, rhythm, and emotion.  

\begin{table} 
\begin{center}
\begin{tabular}{l|r|r}
\hline
Mtxl vs DeepJ (30 sec.) & Mtxl & DeepJ \\ 
\hline
Professional group & \textbf{64} & 34 \\ 
Overall votes & \textbf{158} & 143 \\ 
\hline\hline
Mtxl vs 1-seq txl (30 sec.) & Mtxl & 1-seq txl \\ 
\hline
Professional group & \textbf{55} & 36 \\ 
Overall votes & \textbf{172} & 129 \\ 
\hline\hline
Mtxl vs MusicT (30 sec.) & Mtxl & MuiscT \\ 
\hline
Professional group & 29 & \textbf{69} \\ 
Overall votes & 110 & \textbf{163} \\ 
\hline
Mtxl vs MusicT (10 min.) & Mtxl & MuiscT \\ 
\hline
Professional group & \textbf{59} & 39 \\ 
Overall votes & \textbf{124} & 72 \\ 
\hline
\end{tabular}
\end{center}
\caption{Scores of votes for the four comparisons. }\label{table_human_eva_res}
\end{table}

The results are shown in Table \ref{table_human_eva_res}. Our music significantly exceeds ($p<0.01$) the ones generated by DeepJ, 1-seq txl, and 10-minute music of MusicT. People prefer the music generated by Music Transformer the most only at the case of 30-second music. The major reason collected from composer evaluators is that, the more ``humanistic'' music generated by Music Transformer is due to their unstable tempo and extremely long notes which make the music feel impressionism and have a natural reverb sounds similar to adding the sustain pedal. 

\begin{figure}
    \centering
    \includegraphics[width=7.8cm]{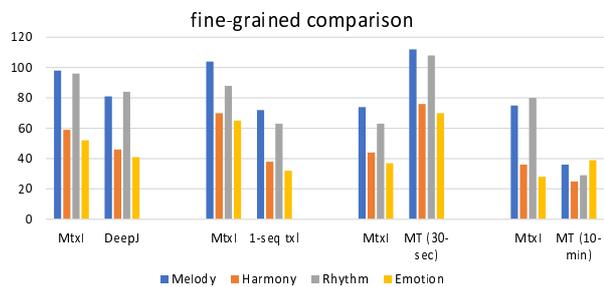}
    \caption{Fine-grained pairwise comparison.}
    \label{figure:why-us}
\end{figure}

Fine-grained comparison is depicted in Figure \ref{figure:why-us}. According to the data we collected from the pairwise comparison test, melody and rhythm are generally weighted higher than harmony and emotion. The results align with Table \ref{table_human_eva_res}: except MusicT of 30-sec., our model achieved significantly better ($p<0.01$) results than DeepJ and 1-seq txl. When we compare Mtxl and 10-min. MusicT, we observe that our melody and rhythm are significantly better ($p<0.01$). There is a tie-situation in terms of harmony and emotion, reflecting requirements of future work on improving these two aspects.

\begin{figure}
    \centering
    \includegraphics[width=7.5cm]{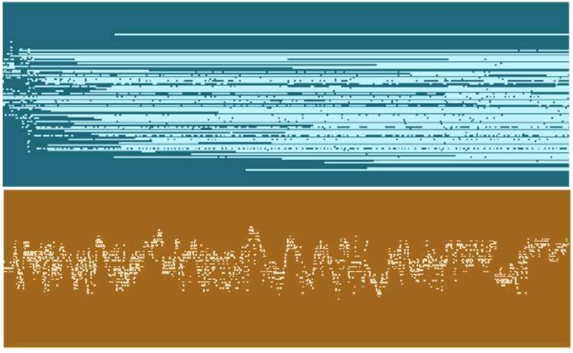}
    \caption{10-minute note sequence comparison of MusicT (up side) and MTxl (bottom side).}
    \label{figure:note-seq-compare}
\end{figure}


For stability evaluation, we pairwise compare the long-time music (10-min.) generated by MusicT and Mtxl. We randomly generated 7 music pieces by each model with the same priming, and extracted the last 30 seconds from each piece. In Table \ref{table_human_eva_res}, our model performs significantly better ($p<0.01$) and is thus more stable in longer-time music generation of MusicT's extremely long notes). For example, in longer-time music, huge number of extremely long notes are generated by Music Transformer but only a few are really attacked and can be heard by human in later stages of music, as illustrated in Figure \ref{figure:note-seq-compare}. 

Inspired by AI generated music, composers can further edit and modify it for specific emotional expression. For instance, Music Transformer and Single-sequence based Transformer-XL generate beaming-sixteenth-like notes yet with unstable tempo, which sounds weird. Therefore, the disadvantages of these two baselines include: due to the fact that notes are not being quantized to grids, it would be more difficult for people to generate music with specific time signatures, and also modify, set MIDI controllers and rearrange the music. In contrast, people can modify the music generated by our model because all our notes have been quantized into time-valued notes properly. More importantly, we can generate music of whatever time signatures we want with our approach. 

\subsection{Objective Evaluations}

\begin{figure}
    \centering
    \subfigure[density comparison (600 seconds)]
    {
    		\begin{minipage}{8cm}
    		\centering
    		\includegraphics[width=7.5cm]{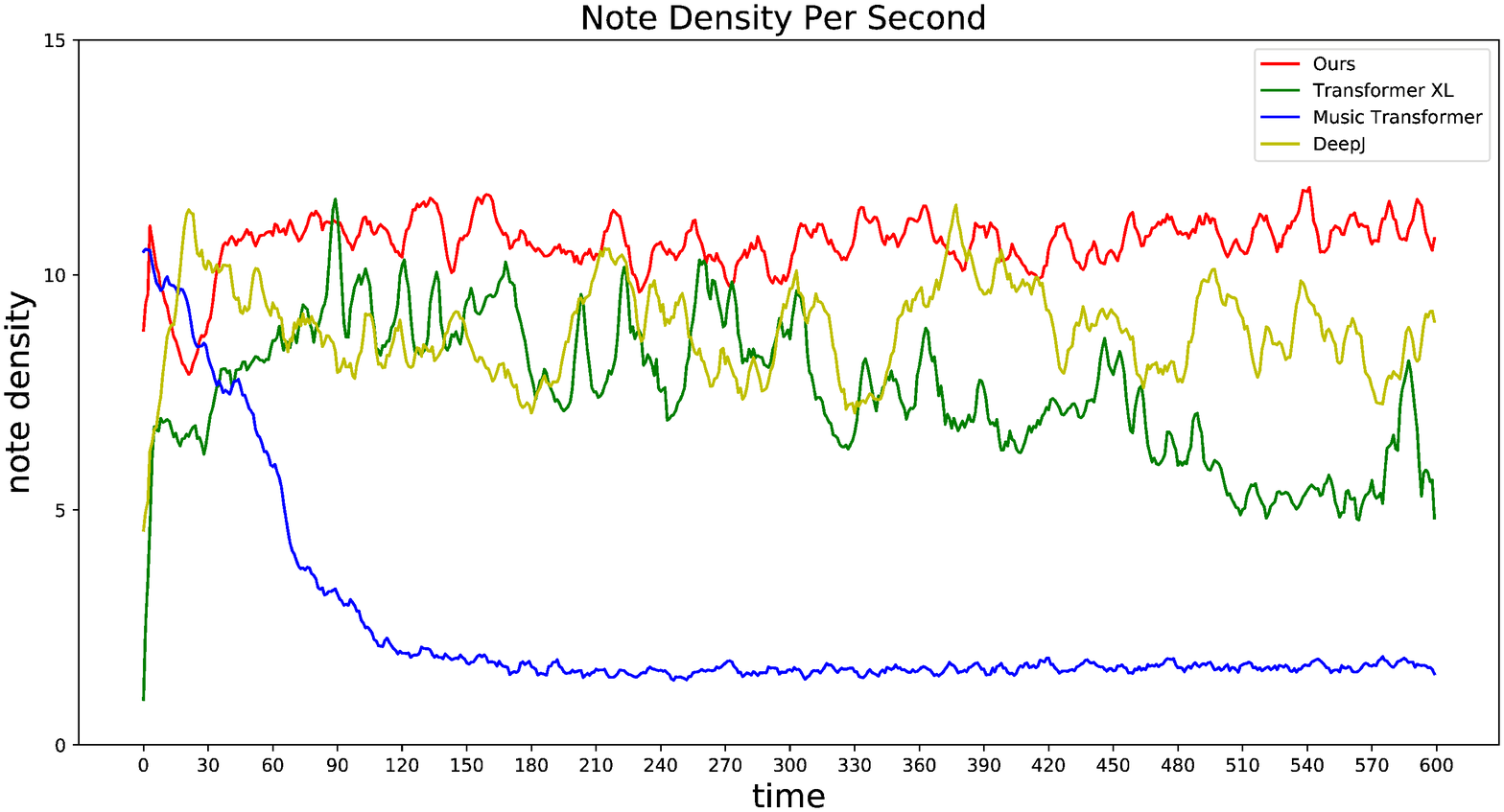}
    		\end{minipage}
    }
    \subfigure[pitch distribution comparison]
    {
    		\begin{minipage}{8cm}
    		\centering
    		\includegraphics[width=7.2cm]{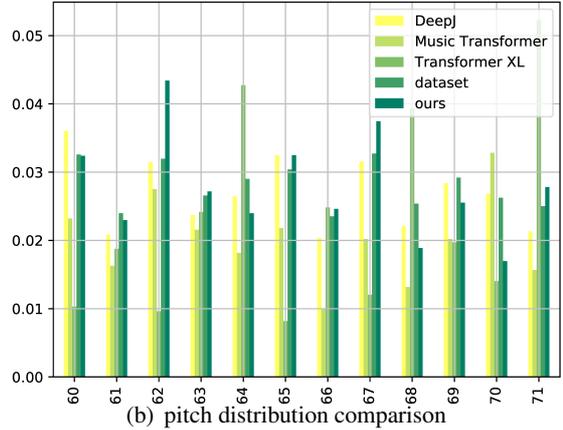}
    		\end{minipage}
    }
    \caption{Density and pitch distribution comparison}
    \label{figure:density-compare}
\end{figure}


We calculate the densities (Figure \ref{figure:density-compare}) of notes using 4$\times$100 samples, 100 from each system. The density is defined to be the number of note-on per window size (5 seconds). As the generated sequences become longer and longer, the densities of Music Transformer and single-sequence Transformer-XL drop seriously. One main reason is the accumulated difficulty of NOTE\_ON and NOTE\_OFF matching for long sequences. As an intuitive comparison, our model can keep its stable note density. In fact, we generated a 36-hour sample, its density keeping to be sustainable. In addition, we compare pitch scattering (Figure \ref{figure:density-compare}) during pitch of 60 to 71. Our model's pitch distribution is the closest to the original dataset, reflecting the strong learning ability of our framework.


\subsection{Some Observations in Music Characteristics}

\subsubsection{Chord-like Progressions}

\begin{figure}
    \centering
    \includegraphics[width=7.8cm]{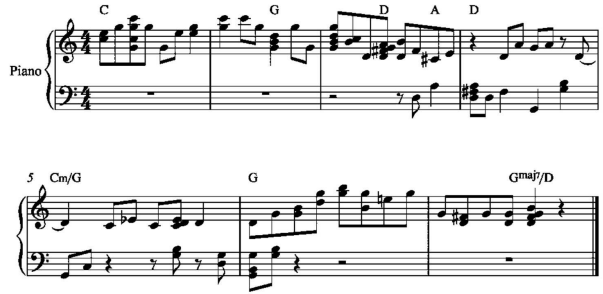}
    \caption{Example of Chord-like progression.}
    \label{figure:example-chord-like-progression}
\end{figure}


As shown in Figure \ref{figure:example-chord-like-progression}, we find that our model has learned some chord progression modes from original data. 
Here, the Tonic-to-Dominant chord progression, one of the most commonly used progressions, has been observed in numerous generated pieces. This successive chord progression also resulted in a tonal transposition from C major to D major. 

\subsubsection{Well-adjusted Dynamics and Rhythm Patterns}

\begin{figure}
    \centering
    \includegraphics[width=7.8cm]{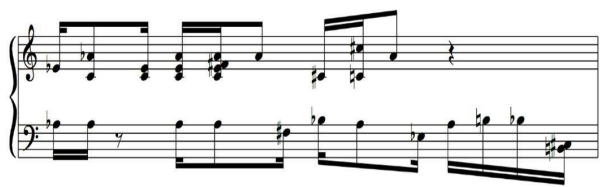}
    \caption{Rhythm patterns example.}
    \label{figure:example-2-rhythm-patterns}
\end{figure}

What surprised us is that the dynamic control by ``independent'' velocity performed well in our results. Few notes have abrupt velocity such as a sudden high or a sudden low. The velocity was mostly changing gradually with a tender ascending and descending. Also, between sections, there are some whole-group contrasts in velocity which are quite similar to the emotional contrast between music movements. Moreover, the velocity ascends and descends along with the rising pitches and the falls, commonly used in real music compositions. In a quarter note-long example (Figure \ref{figure:example-2-rhythm-patterns}\footnote{{https://www.youtube.com/watch?v=V8XSCojvaas}}), besides some common combinations such as single quarter note, beaming eighth notes, one eighth plus two sixteenth, beaming sixteenth notes, there are syncopated patterns, notes with dots and ties and appropriate rests. 

\section{Conclusion}\label{section:conclusion}

We have described our Transformer-XL based piano music generation framework and experiments. Motived by training from composers' music book scores instead of players' performance events, we leverage four sequences of time-valued notes, former note on to current note on, current note on to its note off, pitch, and velocity. We first proposed this novel note representation method which projects each note in MIDI files into a 4-tuple. Then, four Transformer-XL networks are jointly trained by taking these four sequences as inputs with shared embedding concatenations and accumulated cross entropy loss. Through training on a 23-hour piano MIDI dataset, our framework generates significantly better and hour-level stably longer music than three state-of-the-art baselines, Music Transformer, DeepJ, and single sequence-based Transformer-XL, evaluated automatically and manually.

Our multi-sequence learning framework for music is scalable and can be further enriched by additional information, such as tempo sequences, emotions and music genres for enhancing the expressive ability of the generated music. We take these as our future work.   

\bibliographystyle{acl_natbib}
\bibliography{ijcai20}

\begin{thebibliography}{17}
\expandafter\ifx\csname natexlab\endcsname\relax\def\natexlab#1{#1}\fi

\bibitem[{Cho et~al.(2014)Cho, van Merri{\"e}nboer, Gulcehre, Bahdanau,
  Bougares, Schwenk, and Bengio}]{cho-etal-2014-learning}
Kyunghyun Cho, Bart van Merri{\"e}nboer, Caglar Gulcehre, Dzmitry Bahdanau,
  Fethi Bougares, Holger Schwenk, and Yoshua Bengio. 2014.
\newblock Learning phrase representations using {RNN} encoder{--}decoder for
  statistical machine translation.
\newblock In \emph{Proceedings of the 2014 Conference on Empirical Methods in
  Natural Language Processing ({EMNLP})}, pages 1724--1734, Doha, Qatar.
  Association for Computational Linguistics.

\bibitem[{Dai et~al.(2019)Dai, Yang, Yang, Carbonell, Le, and
  Salakhutdinov}]{dai-etal-2019-transformer}
Zihang Dai, Zhilin Yang, Yiming Yang, Jaime Carbonell, Quoc Le, and Ruslan
  Salakhutdinov. 2019.
\newblock Transformer-{XL}: Attentive language models beyond a fixed-length
  context.
\newblock In \emph{Proceedings of the 57th Annual Meeting of the Association
  for Computational Linguistics}, pages 2978--2988, Florence, Italy.
  Association for Computational Linguistics.

\bibitem[{Devlin et~al.(2019)Devlin, Chang, Lee, and
  Toutanova}]{devlin-etal-2019-bert}
Jacob Devlin, Ming-Wei Chang, Kenton Lee, and Kristina Toutanova. 2019.
\newblock {BERT}: Pre-training of deep bidirectional transformers for language
  understanding.
\newblock In \emph{Proceedings of the 2019 Conference of the North {A}merican
  Chapter of the Association for Computational Linguistics: Human Language
  Technologies, Volume 1 (Long and Short Papers)}, pages 4171--4186,
  Minneapolis, Minnesota. Association for Computational Linguistics.

\bibitem[{Donahue et~al.(2019)Donahue, Mao, Li, Cottrell, and
  McAuley}]{donahue2019lakhnes}
Chris Donahue, Huanru~Henry Mao, Yiting~Ethan Li, Garrison~W. Cottrell, and
  Julian McAuley. 2019.
\newblock Lakhnes: Improving multi-instrumental music generation with
  cross-domain pre-training.
\newblock In \emph{ISMIR}.

\bibitem[{Hadjeres and Pachet(2017)}]{deepbach2017}
Ga{\"{e}}tan Hadjeres and Fran{\c{c}}ois Pachet. 2017.
\newblock Deepbach: a steerable model for bach chorales generation.
\newblock In \emph{ICML}.

\bibitem[{Hawthorne et~al.(2018)Hawthorne, Huang, Ippolito, and
  Eck}]{transformer-nade-2018}
Curtis Hawthorne, Anna Huang, Daphne Ippolito, and Douglas Eck. 2018.
\newblock \href {https://nips2018creativity.github.io/doc/Transformer_NADE.pdf}
  {Transformer-nade for piano performances}.
\newblock In \emph{Advances in Neural Information Processing Systems}.

\bibitem[{Hochreiter and Schmidhuber(1997)}]{hochreiter1997long}
Sepp Hochreiter and J{\"u}rgen Schmidhuber. 1997.
\newblock \href
  {https://www.mitpressjournals.org/doi/10.1162/neco.1997.9.8.1735} {Long
  short-term memory}.
\newblock \emph{Neural computation}, 9(8):1735--1780.

\bibitem[{Huang et~al.(2018)Huang, Vaswani, Uszkoreit, Shazeer, Hawthorne, Dai,
  Hoffman, and Eck}]{music-transformer-2018}
Cheng{-}Zhi~Anna Huang, Ashish Vaswani, Jakob Uszkoreit, Noam Shazeer, Curtis
  Hawthorne, Andrew~M. Dai, Matthew~D. Hoffman, and Douglas Eck. 2018.
\newblock An improved relative self-attention mechanism for transformer with
  application to music generation.
\newblock \emph{CoRR}.

\bibitem[{Johnson(2017)}]{biaxial-lstm2017}
Daniel Johnson. 2017.
\newblock \href {https://doi.org/10.1007/978-3-319-55750-2_9} {\emph{Generating
  Polyphonic Music Using Tied Parallel Networks}}, volume 10198 of \emph{In:
  Correia J., Ciesielski V., Liapis A. (eds) Computational Intelligence in
  Music, Sound, Art and Design. EvoMUSART 2017. Lecture Notes in Computer
  Science}.
\newblock Springer, Cham.

\bibitem[{Kingma and Ba(2014)}]{kingma2014method}
Diederik~P. Kingma and Jimmy Ba. 2014.
\newblock \href {http://arxiv.org/abs/1412.6980} {Adam: A method for stochastic
  optimization}.
\newblock Cite arxiv:1412.6980 Comment: Published as a conference paper at the
  3rd International Conference for Learning Representations, San Diego, 2015.

\bibitem[{Li et~al.(2019)Li, Liu, Liu, Zhao, Liu, and
  Zhou}]{DBLP:journals/corr/abs-1809-08895}
Naihan Li, Shujie Liu, Yanqing Liu, Sheng Zhao, Ming Liu, and Ming Zhou. 2019.
\newblock \href {http://arxiv.org/abs/1809.08895} {Neural speech synthesis with
  transformer network}.
\newblock In \emph{Proceedings of the AAAI Conference on Artificial
  Intelligence}, volume abs/1809.08895.

\bibitem[{Mao et~al.(2018)Mao, Shin, and Cottrell}]{deepj2018}
Huanru~Henry Mao, Taylor Shin, and Garrison~W. Cottrell. 2018.
\newblock \href {http://arxiv.org/abs/1801.00887} {Deepj: Style-specific music
  generation}.
\newblock In \emph{2018 IEEE 12th International Conference on Semantic
  Computing (ICSC)}, volume abs/1801.00887.

\bibitem[{Mohamed et~al.(2019)Mohamed, Okhonko, and
  Zettlemoyer}]{DBLP:journals/corr/abs-1904-11660}
Abdelrahman Mohamed, Dmytro Okhonko, and Luke Zettlemoyer. 2019.
\newblock \href {http://arxiv.org/abs/1904.11660} {Transformers with
  convolutional context for {ASR}}.
\newblock \emph{CoRR}, abs/1904.11660.

\bibitem[{Raffel and Ellis(2014)}]{pretty-midi2014}
Colin Raffel and Daniel P.~W. Ellis. 2014.
\newblock Intuitive analysis, creation and manipulation of midi data with
  pretty midi.
\newblock In \emph{15th International Conference on Music Information Retrieval
  Late Breaking and Demo Papers}.

\bibitem[{Shaw et~al.(2018)Shaw, Uszkoreit, and Vaswani}]{shaw-etal-2018-self}
Peter Shaw, Jakob Uszkoreit, and Ashish Vaswani. 2018.
\newblock \href {https://doi.org/10.18653/v1/N18-2074} {Self-attention with
  relative position representations}.
\newblock In \emph{Proceedings of the 2018 Conference of the North {A}merican
  Chapter of the Association for Computational Linguistics: Human Language
  Technologies, Volume 2 (Short Papers)}, pages 464--468, New Orleans,
  Louisiana. Association for Computational Linguistics.

\bibitem[{Vaswani et~al.(2017)Vaswani, Shazeer, Parmar, Uszkoreit, Jones,
  Gomez, Kaiser, and Polosukhin}]{NIPS2017_7181_transformer}
Ashish Vaswani, Noam Shazeer, Niki Parmar, Jakob Uszkoreit, Llion Jones,
  Aidan~N Gomez, \L~ukasz Kaiser, and Illia Polosukhin. 2017.
\newblock \href
  {http://papers.nips.cc/paper/7181-attention-is-all-you-need.pdf} {Attention
  is all you need}.
\newblock In I.~Guyon, U.~V. Luxburg, S.~Bengio, H.~Wallach, R.~Fergus,
  S.~Vishwanathan, and R.~Garnett, editors, \emph{Advances in Neural
  Information Processing Systems 30}, pages 5998--6008. Curran Associates, Inc.

\bibitem[{Zhu et~al.(2018)Zhu, Chen, Liu, Yuan, Qin, Li, Zhang, Zhou, Wei, and
  Xu}]{xiaoiceband2018wxc}
Hongyuan Zhu, Enhong Chen, Qi~Liu, Nicholas Yuan, Chuan Qin, Jiawei Li, Kun
  Zhang, Guang Zhou, Furu Wei, and Yuanchun Xu. 2018.
\newblock Xiaoice band: A melody and arrangement generation framework for pop
  music.
\newblock In \emph{Conference: the 24th ACM SIGKDD International Conference},
  pages 2837--2846.

\end{thebibliography}

\end{document}